\documentclass[%
reprint,
 superscriptaddress,
 floatfix,
 amsmath,amssymb,
 pra,
]{revtex4-1}

\usepackage{graphicx}% Include figure files
\usepackage{dcolumn}% Align table columns on decimal point
\usepackage{bm}% bold math
\usepackage{amsmath}
\usepackage{hyperref}% add hypertext capabilities
\usepackage{lipsum}

\usepackage{tabularx}

\begin{document}

\title{Lifetime Measurements of the $A^2\Pi_{1/2}$ and $A^2\Pi_{3/2}$ States in BaF}

\author{P.~Aggarwal}
\author{V.R.~Marshall}
\affiliation{Van Swinderen Institute for Particle Physics and Gravity, University of Groningen, The Netherlands}
\affiliation{Nikhef, National Institute for Subatomic Physics, Amsterdam, The Netherlands}

\author{H.L.~Bethlem}
\affiliation{Van Swinderen Institute for Particle Physics and Gravity, University of Groningen, The Netherlands}
\affiliation{Department of Physics and Astronomy, and LaserLaB, Vrije Universiteit Amsterdam, The Netherlands}
\author{A.~Boeschoten}
\author{A.~Borschevsky}
\author{M.~Denis}
\author{K.~Esajas}
\author{Y.~Hao} 
\affiliation{Van Swinderen Institute for Particle Physics and Gravity, University of Groningen, The Netherlands}
\affiliation{Nikhef, National Institute for Subatomic Physics, Amsterdam, The Netherlands}
\author{S.~Hoekstra}
 \email{s.hoekstra@rug.nl}
\affiliation{Van Swinderen Institute for Particle Physics and Gravity, University of Groningen, The Netherlands}
\affiliation{Nikhef, National Institute for Subatomic Physics, Amsterdam, The Netherlands}
\author{K.~Jungmann}
\author{T.B.~Meijknecht}
\affiliation{Van Swinderen Institute for Particle Physics and Gravity, University of Groningen, The Netherlands}
\affiliation{Nikhef, National Institute for Subatomic Physics, Amsterdam, The Netherlands}
\author{M.C.~Mooij}
\affiliation{Nikhef, National Institute for Subatomic Physics, Amsterdam, The Netherlands}
\affiliation{Department of Physics and Astronomy, and LaserLaB, Vrije Universiteit Amsterdam, The Netherlands}
\author{R.G.E.~Timmermans}
\author{A.~Touwen}
\affiliation{Van Swinderen Institute for Particle Physics and Gravity, University of Groningen, The Netherlands}
\affiliation{Nikhef, National Institute for Subatomic Physics, Amsterdam, The Netherlands}
\author{W.~Ubachs}
\affiliation{Department of Physics and Astronomy, and LaserLaB, Vrije Universiteit Amsterdam, The Netherlands}
\author{S.M.~Vermeulen}
\author{L.~Willmann}
\author{Y.~Yin}
\author{A.~Zapara}
\affiliation{Van Swinderen Institute for Particle Physics and Gravity, University of Groningen, The Netherlands}
\affiliation{Nikhef, National Institute for Subatomic Physics, Amsterdam, The Netherlands}

\collaboration{NL-eEDM Collaboration}%\noaffiliation

\date{\today}% It is always \today, today,
             %  but any date may be explicitly specified

\begin{abstract}
{
Time resolved detection of laser induced fluorescence from pulsed excitation of electronic states in barium monofluoride (BaF) molecules has been performed in order to determine the lifetimes of the $A^2\Pi_{1/2}$ and $A^2\Pi_{3/2}$ states. The method permits control over experimental parameters such that systematic biases in the interpretation of the data can be controlled to below $10^{-3}$ relative accuracy. The statistically limited values for the lifetimes of the $A^2\Pi_{1/2}(\nu=0)$ and $A^2\Pi_{3/2}(\nu=0)$ states are 57.1(3) ns and 47.9(7)~ns, respectively. The ratio of these values is in good agreement with scaling for the different excitation energies. The investigated molecular states are of relevance for an experimental search for a permanent electric dipole moment (EDM) of the electron in BaF.
}
\end{abstract}

%\pacs{PACS need to be selected}
\maketitle

Molecular systems, including barium monofluoride (BaF), have received significant attention in the context of experimental investigations of nuclear-spin-dependent parity violation effects \cite{Altunta2018} and searches for EDMs \cite{Hudson2011, Cairncross2017, Vutha2018, Andreev2018,  Aggarwal2018}. For this reason, activity has commenced on molecular structure calculations of BaF \cite{Kang2016, Chen2016s, Tohme2015, Hao2019}. A relevant and experimentally accessible property is the radiative lifetime of electronically excited states. The measurement of lifetimes in cesium~\cite{Young1994, Toh2018} for example, in combination with atomic structure calculations, has impacted the interpretation of atomic-parity violation~\cite{Safronova2018}.

Previous measurements on radiative lifetimes in the BaF molecule have been performed in a resistance furnace using pulsed dye laser light \cite{Dagdigian1974,Berg1993,Berg1998}. These measurements were limited by contributions from collisions and background which are not related to the time-dependent fluorescence from molecules. The control over pulse length and timing accuracy of these early measurements can be improved due to technology available today.

We demonstrate a nearly background-free method for the determination of excited state radiative lifetimes, which are sensitive to the entire spatial extent of the wavefunctions. In particular, we investigate the $A^2\Pi_{1/2}(\nu=0)$ and $A^2\Pi_{3/2}(\nu=0)$ states in BaF, which govern the possibility of laser cooling of this molecular system \cite{Kang2016, Chen2016s, Xu2017, Hao2019}. 

The measurement is performed on a fast internally cold supersonic molecular beam in vacuum conditions. The molecules are excited by short pulses of resonant laser radiation. Laser-induced fluorescence (LIF) is detected by a photomultiplier tube (PMT) with sub-nanosecond time resolution with respect to the excitation pulse. The accuracy of the timing for the laser pulses and the detection is derived from a GPS-stabilized reference clock. A detailed study of systematic biases is possible in our experiment due to a low rate of background photons in the detection systems and due to simultaneous measurement of the PMT response. The analysis shows that these biases can be controlled to a level of $10^{-3}$ on the yielded lifetimes.

%\section{\label{sec:Methods}Method}

\begin{figure}
    \includegraphics[scale=1,width=1\linewidth]{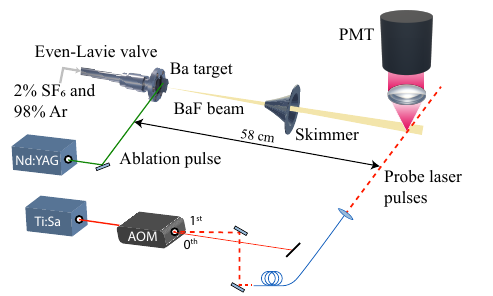}
    \caption{Schematic overview of the experimental setup. BaF molecules are produced in a supersonic source and pass through a skimmer into the interaction region. Pulses are generated from the CW light of a Ti-Sapphire (Ti:Sa) laser using the first-order diffracted beam from an acousto-optic modulator (AOM). The light from the Ti:Sa laser is delivered to the molecular beam via a 50~m single-mode optical fibre.}
    \label{fig:setup}
\end{figure}

Metal fluoride molecules, such as BaF, can be produced efficiently in a supersonic beam source \cite{ Tarbutt2002, Steimle2011, Mathavan2016, Cournol2018}. The molecules are excited to the higher electronic state by short laser pulses (see Fig.~\ref{fig:setup}). The mean lifetime value of the molecules in the excited state is extracted from the fluorescence decay after the laser pulses.

The supersonic ablation source operates at a repetition rate of 10~Hz. An Even-Lavie valve \cite{Even2014} creates short ($\sim 35~\mu$s) pulses of carrier gas, which is a mixture of 98~\% argon (Ar) and 2~$\%$ sulphur hexafluoride (SF$_6$). The disc-shaped barium metal target is mounted inside the vacuum chamber at a distance of 6~mm perpendicular to the supersonic beam. Barium atoms are ablated by 10~ns long pulses from a Nd:YAG laser, with an energy of 20~mJ at a wavelength of 532~nm. The ablation products are entrained in the carrier gas and react with $\rm{SF_6}$ to form BaF molecules which cool in the supersonic expansion. The molecular beam then passes through a skimmer of diameter 4~mm, located 28~cm downstream of the valve, and enters a second chamber where the molecules are detected by LIF. 

The detection laser light is generated from the continuous wave (CW) output of a Ti-Sapphire laser. Pulses are created from this light by selecting the first order diffracted beam from an acousto-optic modulator (AOM, Brimrose TEM-200-50) powered by radio frequency (RF) at 200~MHz. The pulse structure and amplitude are controlled by switching the RF (Mini-Circuits ZASWA-2-50-DR+) to obtain pulses which are synchronized to interact with the passing molecular pulse. The fraction of the first order diffracted light beam in the fundamental Gaussian mode is coupled into a 50~m long single-mode optical fibre, before sending it to the supersonic beam setup. 

The laser pulses intersect with the molecular beam perpendicularly, exciting molecules on the $A^{2}{\Pi_{1/2}}(\rm{\nu}=~0,~J=1/2)~\leftarrow~X^2\Sigma^{+}({\nu}=0)$ (860~nm) or $A^{2}{\Pi_{3/2}}(\rm{\nu}=~0,~J=3/2)~\leftarrow~X^2\Sigma^{+}({\nu}=0)$ (815~nm) electronic transition. The fluorescence emitted is collected by a 2~inch diameter achromatic lens with a focal length of 75~mm, which focuses the light onto an infrared-sensitive PMT (H7422-50 Hamamatsu). A narrow bandpass filter with 60~$\%$ transmission at the resonance wavelength selects fluorescence photons. The detection chamber is internally blackened with paint that absorbs infrared light (AZ Technology MLS-85-SB) in order to reduce the background count rate. The cooled PMT has a count rate of about 30~s$^{-1}$ from thermionic emission and of 2000~s$^{-1}$ for CW laser light of power 1~mW passing through the interaction zone. 

\begin{figure}[tb]
    \includegraphics[scale=1,width=1\linewidth]{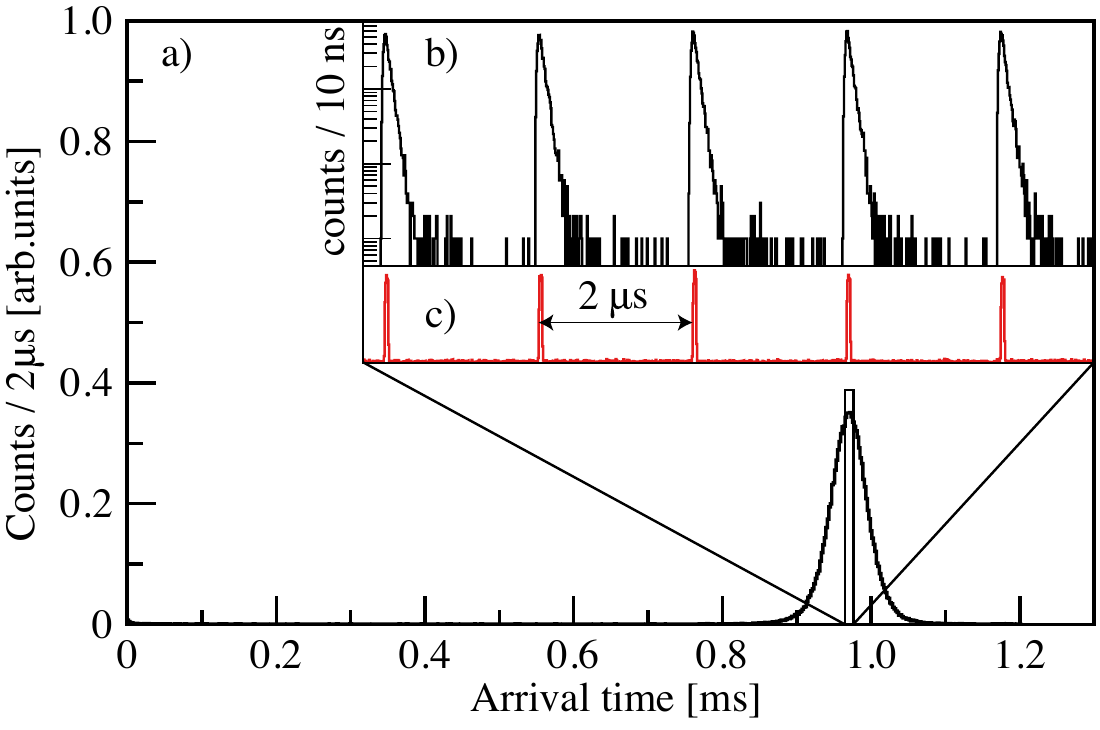}
    \caption{Time-of-flight profile showing the accumulated fluorescence of the BaF molecular beam following excitation on the $A\leftarrow X$ transition. The histogram in panel a) shows the arrival-time distribution of the molecular beam fluorescence, binned at 2 $\mu$s. Panel b) shows the decay of the fluorescence following excitation by 40~ns laser pulses (shown in panel c)) at repetition rate of 500 kHz, reflecting the lifetime of the A$^{2}\Pi_{1/2}$ state.}
    \label{fig:TOF}
\end{figure}

A pulse train is generated with a period of 2~$\mu$s, a duty cycle of 2~\%, and a pulse length of 40~ns. Since the duration of the excitation pulse is comparable to the lifetime of the excited state, the Fourier-limited bandwidth of the pulses is comparable to the linewidth of the employed transitions. 

The $100$~ms time period between two molecular pulses is split into $50$~ms with the pulse train and $50$~ms without laser light. The former permits the determination of the laser pulse parameters, while the latter determines the dark count rate of the PMT. The molecules arrive at the interaction region around $1$~ms after the laser ablation pulse (Fig.~\ref{fig:TOF}a). On timescales of tens of ns the response to excitation laser pulses (Fig.~\ref{fig:TOF}c) and the molecular fluoresence (Fig.~\ref{fig:TOF}b) is observed. The arrival time of each photon is recorded by a multihit-capable time to digital converter (Signadyne, SD-TDC-H3344-PXIe). Accurate timing synchronization is achieved by referencing all timing units to a GPS disciplined $10$~MHz Rubidium Frequency Standard ($\Delta \nu$ = $10^{-12}$ in $2000$~s, FS725 Stanford Research Systems).  

The molecular beam has a velocity of ($594\pm18$)~ms$^{-1}$ corresponding to a travel of 60~$\mu$m in 100~ns.Therefore, the molecules do not travel significantly in or out of the detection zone during the measurement. Up to $100$~photons per ablation shot are detected with CW detection laser light with a power of $2$~mW and spot size of $4$~mm. This is comparable to the molecular beam diameter size.

The timing parameters of the laser pulses are measured simultaneously with the same data acquisition system as the molecular signal. The fall time (between 90 and 10~$\%$ of the amplitude) of the laser pulses is additionally measured with a fast photodiode (2~GHz bandwidth) on an oscilloscope to be $11(1)$~ns. The ratio of the light intensity during the laser pulse to the intensity in the following several hundred ns, is determined to be larger than 1000:1. 

\begin{figure}[tb]
    \includegraphics[scale=1,width=1.\linewidth]{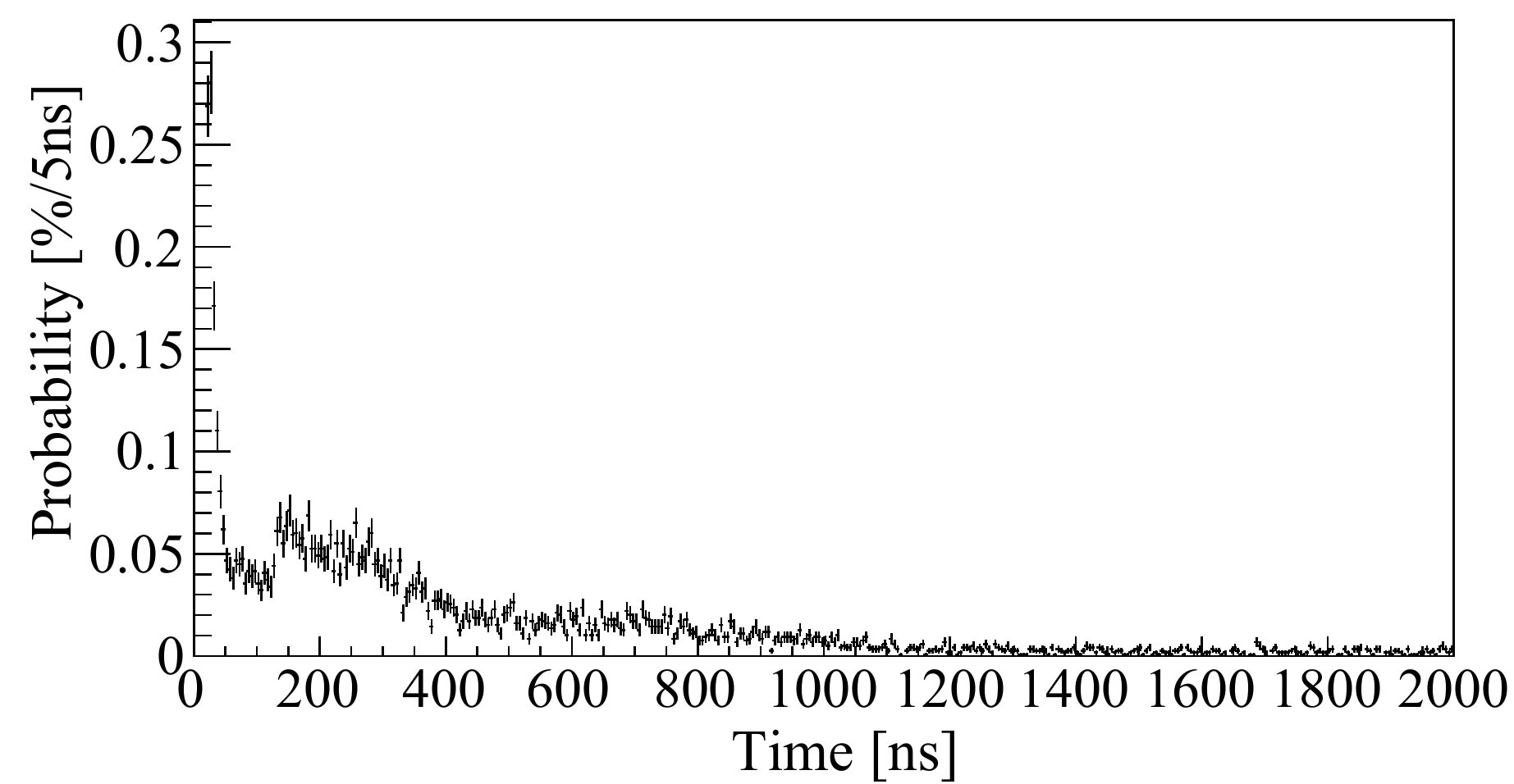}
    \caption{Time distribution of the probability of a second photon in a time window of 2000~ns. This spectrum is extracted from the same dataset as the fluorescence data for the determination of state lifetimes. } 
     \label{fig:afterpulses}
\end{figure}

%\section{Results}
In single-photon counting with PMTs, each pulse has $\sim$1\% probability to be followed by a second pulse (afterpulse)~\cite{Burstyn1980,Zhao2003, Ishii2015}. These afterpulses cause the main systematic bias in our low-background experiment. The employed multihit-capable data acquisition system permits quantitative recording of the time distribution of these pulses (see Fig.~\ref{fig:afterpulses}) throughout all data taking periods for the lifetime measurements. The distribution of these pulses ranges over timescales of $\mu$s with a structure comparable to the LIF signal. This contribution, although small, is taken into account in the determination of the excited state lifetimes. 

The time distribution $N(t)$ of the number of photons recorded at time $t$ after the trailing edge of the excitation laser pulse is described by

\begin{equation}
N(t) =  \frac{A}{\tau} e^{-t/\tau} + B(t) + C,
\label{equation:singleexp}
\end{equation}   
where $A$ is the number of photons recorded in the exponential decay, $\tau$ is the lifetime of the excited state, $B(t)$ is the convolution of the distribution of afterpulses (Fig. \ref{fig:afterpulses}) with the distribution $N(t)$, and $C$ is a constant offset due to background counts. The uncertainties on the counts per channel arise from Poissonian statistics.

\begin{figure}[htb]
    \includegraphics[scale=1,width=1\linewidth]{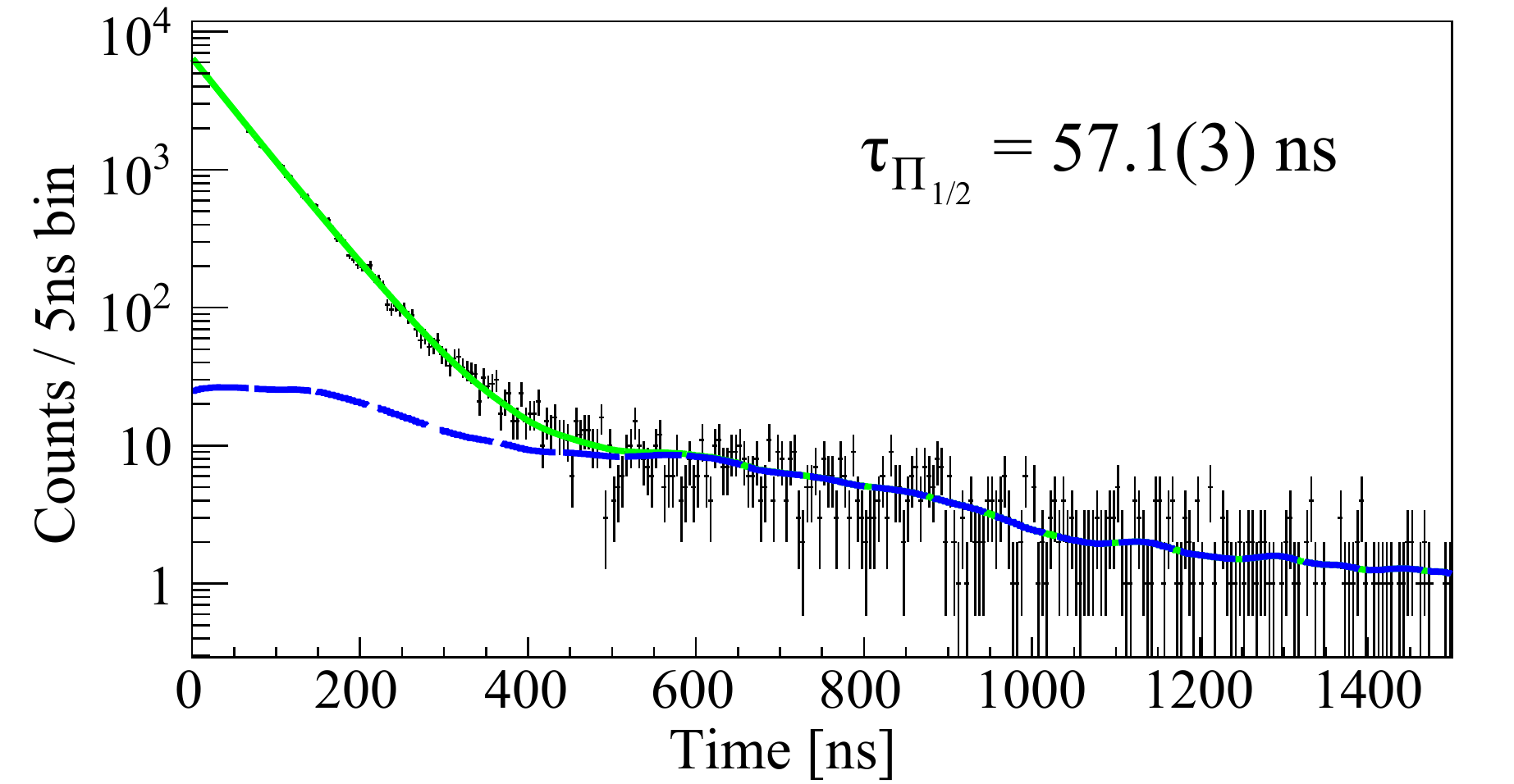}
    \caption{Accumulated spectrum for the time distribution $N(t)$ for the $A^{2}{\Pi_{1/2}}(\nu=0)$ state. A maximum-likelihood fit of Eq. \ref{equation:singleexp} to the data is shown by the solid (green) line. The dashed (blue) line indicates the contribution of the time-dependent background $B(t)$. A value for the lifetime of $\tau_{\Pi_{1/2}}$ = 57.1(3) ns is obtained.}
     \label{fig:Pi12_all}
\end{figure}

A maximum likelihood fitting procedure of the function given in Eq. \ref{equation:singleexp} to the experimental data yields a value of $\tau_{\Pi_{1/2}}$ = 57.1(3)~ns for the lifetime (Fig.~\ref{fig:Pi12_all}). The contribution of the time dependent term $B(t)$ is 1.6~\% of the total  statistics. Not taking $B(t)$ into account leads to a bias of 1.6~ns
on the lifetime, which is five times larger than the statistical uncertainty.

\begin{figure}[htb]
    \includegraphics[scale=1,width=1\linewidth]{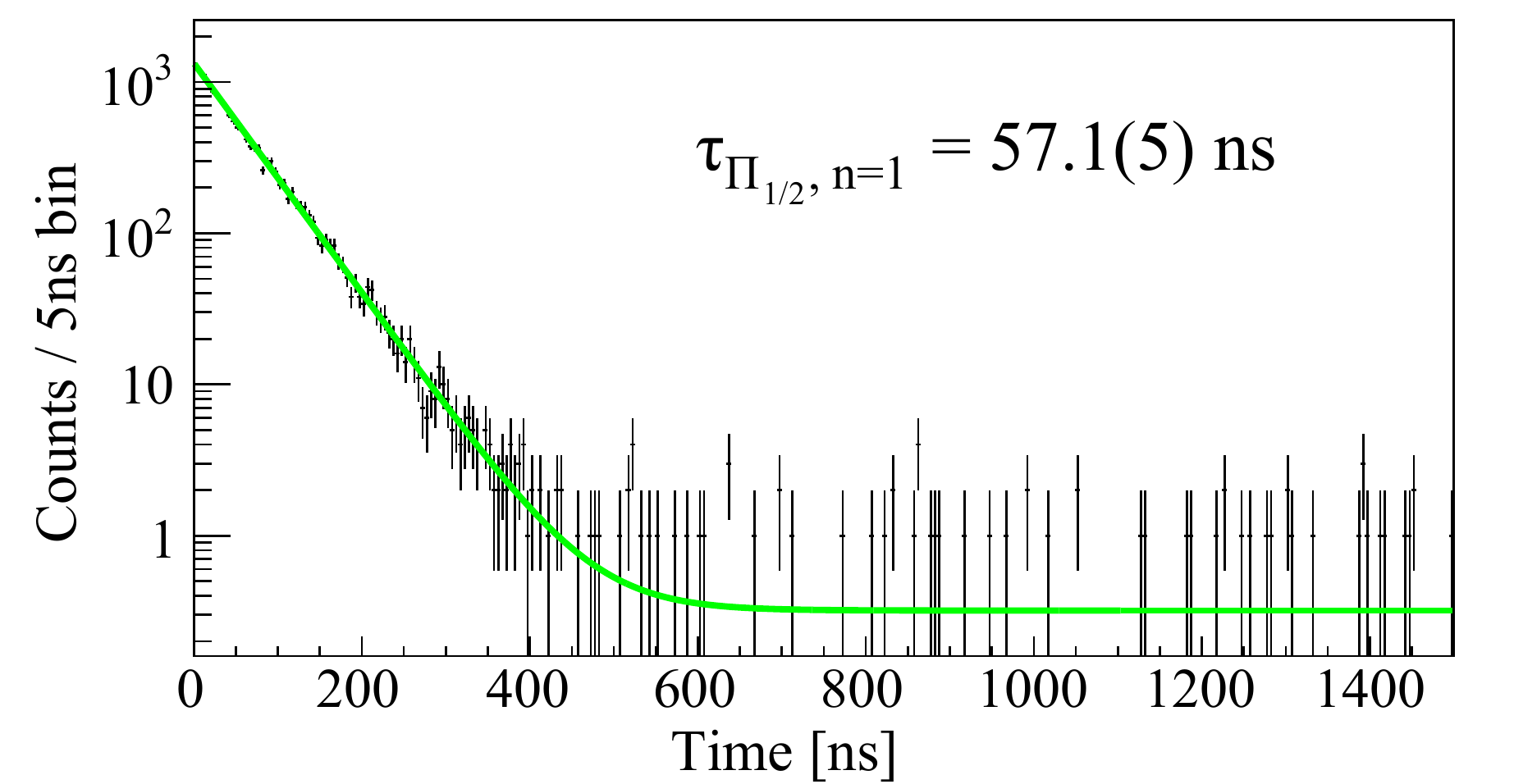}
    \caption{Accumulated spectrum of the time distribution $N(t)$ for the $A^{2}{\Pi_{1/2}}$ state, selecting events with one photon per molecular pulse. The yielded lifetime $\tau_{\Pi_{1/2},n=1}$ = 57.1(5)~ns. }
     \label{fig:Pi12_one_photon}
\end{figure}

\begin{figure}[htb]
    \includegraphics[scale=1,width=1\linewidth]{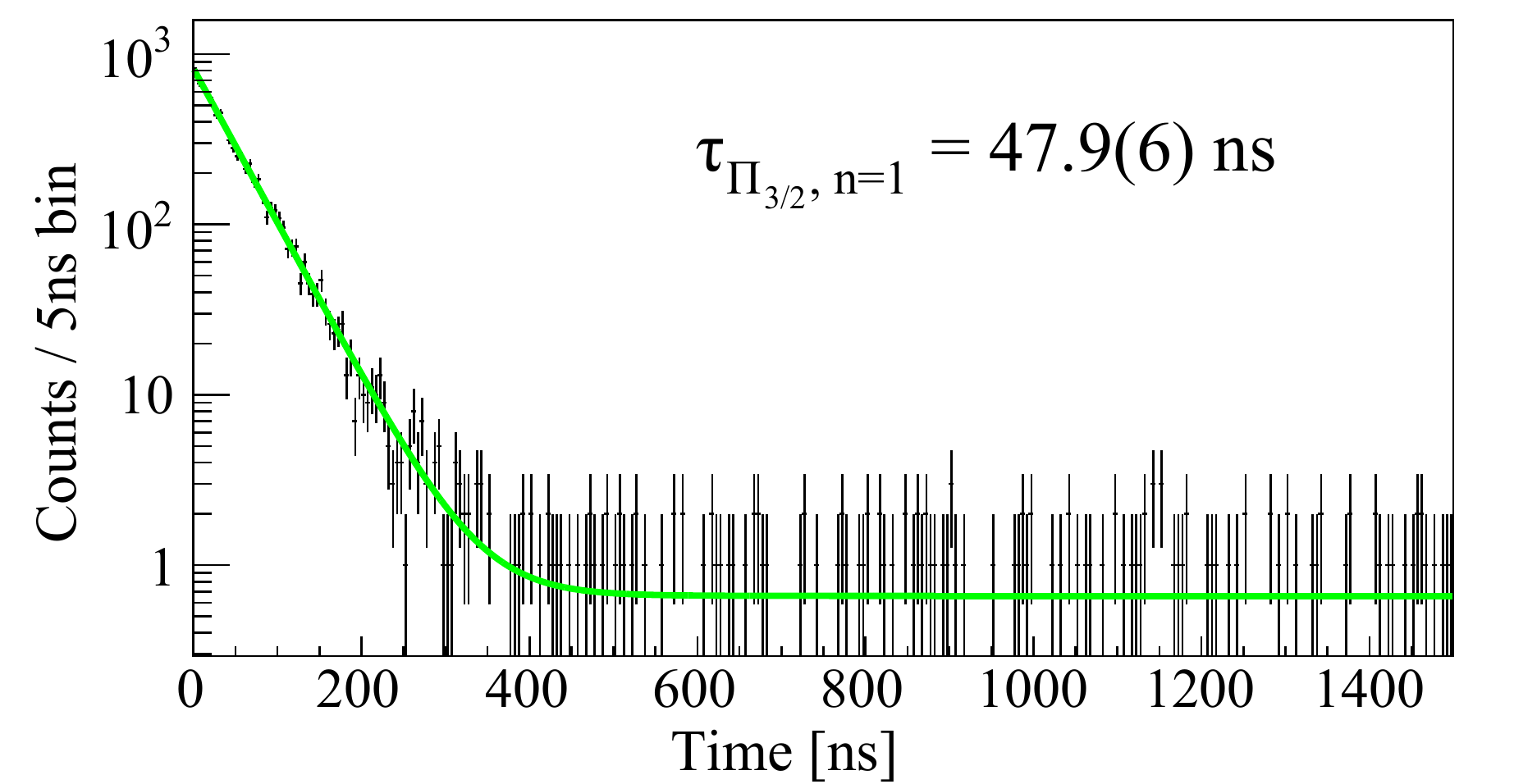}
    \caption{Accumulated spectrum of the time distribution $N(t)$ for the $A^{2}{\Pi_{3/2}}$ state. Selecting events with one photon detected per molecular pulse yielded for the lifetime $\tau_{\Pi_{3/2},n=1} = 47.9(6)$~ns.}
     \label{fig:P32_one_photon}
\end{figure}

On average, three photons per molecular pulse ($\bar n =3$) were recorded. For a quarter of these molecular pulses, exactly 1 photon ($n=1$) was recorded. In this subset of the data, there is no contribution from afterpulses. The resulting spectrum is therefore analyzed with the model function Eq.~\ref{equation:singleexp} with the contribution $B(t)$ set to zero (see Fig.~\ref{fig:Pi12_one_photon}). For the $A^{2}{\Pi_{1/2}}$ state, the lifetime $\tau_{\Pi_{1/2}, n=1}$ = 57.1(5)~ns. This number is in excellent agreement with the result obtained with the full dataset. The same analysis method applied to the recorded time spectra for the $A^{2}{\Pi_{3/2}}(\nu=0)$ state yields the lifetime $\tau_{\Pi_{3/2}, n=1}$ = 47.9(6)~ns (see Fig.~\ref{fig:P32_one_photon}).

The magnetic field in the measurement region has been determined to be $< 50 \mu$T. It is directed vertically, coinciding with the observation line of the PMT. The excitation laser light was elliptically polarized. In such a geometry, modulation of the fluorescence from the excited state due to Zeeman quantum beats~\cite{Young1994} is strongly suppressed. The estimated periods of the quantum beats are 0.9 $\mu$s and 3.3 $\mu$s for the $A^2\Pi_{3/2}$ and the $A^2\Pi_{1/2}$ state, respectively.  

Systematic bias on the measurement due to the excitation light pulse parameters was checked by variation of the duty cycle between 1~$\%$ and 20~$\%$ and of the repetition period between 1~$\mu$s and 10~$\mu$s.
From a systematic scan of the start time for the fitting procedure in Fig.~\ref{fig:Pi12_all}, we find that the variation of the lifetime value is one order of magnitude smaller than the statistical uncertainty. Similar size contributions are observed for subsets of the data with different excitation probabilities due to a change in laser power or drift of the excitation laser frequency.

The measured lifetime values are listed with previous experimental work~\cite{Berg1993,Berg1998} and recent quantum chemistry calculations~\cite{Hao2019, Kang2016} in (Table~\ref{tab:lifetimevalues}). The lifetimes of the excited states of BaF were calculated using the multireference configuration interaction approach in a nonrelativistic framework (NR-MRCI)~\cite{Kang2016}, thus yielding the same result for the $A^{2}{\Pi_{1/2}}$ and the $A^{2}{\Pi_{3/2}}$ states. In a more recent investigation~\cite{Hao2019}, we used relativistic MRCI, where the traditional 4-component  Dirac-Coulomb  (DC)  Hamiltonian  was replaced by  the exact  2-component  Hamiltonian  (X2C-MRCI)  \cite{Iliavs2007, Saue2011}. In both works, experimental level energies were used along with calculated transition dipole moments to determine the lifetimes. The X2C-MRCI results reproduce the lifetime ratio very well, as a result of the accurate treatment of spin-orbit effects. However, the predicted lifetimes underestimate the experimental values by about 30\%, due to the shortcoming of the MRCI approach in the treatment of the transition dipole moments.

\begin{table}[htb]
\begin{tabular}{p{0.55\columnwidth}p{0.2\columnwidth}p{0.2\columnwidth}}
 \hline
 &\multicolumn{2}{c}{Lifetime [ns]}\\

Method & $A^{2}{\Pi_{1/2}}$ & $A^{2}{\Pi_{3/2}}$\\
 \hline
  Supersonic beam (this work)  & 57.1(3) & 47.9(6)\\
  Resistance furnace  & 56.0(9)~\cite{Berg1998} & 46.1(9)~\cite{Berg1993}\\
  X2C-MRCI & 40.4~\cite{Hao2019}  & 34.7~\cite{Hao2019}\\
  NR-MRCI & \multicolumn{2}{c}{37.8~\cite{Kang2016}}\\
  
 \hline
\end{tabular}
\caption{Experimental and theoretical values for the lifetime of the $A^{2}{\Pi_{1/2}}$ and $A^{2}{\Pi_{3/2}}$ states. The theoretical values were obtained in the framework of NR-MRCI and X2C-MRCI methods.} \label{tab:lifetimevalues}
\end{table}

The main decay channel of the investigated excited states is an electric dipole transition to the $X^2\Sigma^{+}({\nu}=0)$ ground state. In this case scaling of lifetimes is expected to be proportional to the third power of the fluorescence wavelength ($\lambda^3$) ($\lambda$ = 859.79~nm and $\lambda$ = 815.45~nm for the $A^{2}{\Pi_{1/2}}$ and $A^{2}{\Pi_{3/2}}$ states, respectively). We compare the ratio of the lifetimes of the two excited states with the previously measured experimental work \cite{Berg1993, Berg1998}, the X2C-MRCI theoretical calculations, and the expected value from $\lambda^3$-scaling. Within the experimental accuracy, our experimental result is in agreement with this scaling (Table~\ref{tab:lifetimeratios}).\\

\begin{table}[t]
\begin{tabular}{p{0.5\columnwidth}p{0.25\columnwidth}}
 \hline
Method & Lifetime ratio \\
 \hline
  This work  & 0.839(14) \\
  $\lambda^3$-scaling & 0.8531 \\
  Resistance furnace~\cite{Berg1993, Berg1998} & 0.823(25) \\
  X2C-MRCI~\cite{Hao2019} & 0.859\\
 \hline
\end{tabular}
\caption{Comparison between the ratios of the lifetimes $\tau_{\Pi_{3/2}} / \tau_{\Pi_{1/2}}$  obtained in this work,  previously measured experimental work \cite{Berg1993, Berg1998}, and the X2C-MRCI theoretical calculations with the expected value from $\lambda^3$-scaling of the transition wavelengths.}
\label{tab:lifetimeratios}
\end{table}

%\section{Conclusion}
A measurement of the lifetimes of the low lying electronically excited $A^2\Pi_{1/2}$ and $A^2\Pi_{3/2}$ states in the BaF molecule was performed in the context of the experimental search for a permanent electric dipole moment of the electron (NL-eEDM)~\cite{Aggarwal2018}. The statistically limited accuracies on the lifetimes of the $A^{2}{\Pi_{1/2}}$ and $A^{2}{\Pi_{3/2}}$ states are 0.5$\%$ and 1.3$\%$, respectively. The main systematic uncertainity arises from the imperfection of photomultipliers in photon counting mode, in particular their afterpulses. By measurement of the afterpulse spectrum of the employed PMT, the relative contributions to the measured lifetime values could be reduced to the 10$^{-3}$ level. Methods which permit identification and control of systematic effects are crucial in precision measurements and their interpretation. Careful design of the experimental procedure and the use of state-of-the-art technology enables the execution of such measurements with low sensitivity to systematic bias.

\section*{Acknowledgements}
The NL-$e$EDM consortium receives program funding (EEDM-166) from the Netherlands Organisation for Scientific Research (NWO). We thank C.J.G Onderwater for discussions on the statistical treatment of the data.

\end{document}